\documentclass[twocolumn,runningheads]{svjour2}
\usepackage{graphicx}
\usepackage[authoryear]{natbib}

\usepackage[latin1]{inputenc}
\usepackage[american]{babel}

\newcommand{\bvp}{{Brunt-Väissälä\ frequency.}}
\newcommand{\cesam}{{\textsc{cesam}\ }}
\newcommand{\cesamp}{{\textsc{cesam.}\ }}
\newcommand{\cesamv}{{\textsc{cesam,}\ }}

\newcommand{\etc}{{\it etc...\ }}
\newcommand{\dsdx}[2]{\frac{\partial #1}{\partial #2}}
\newcommand{\ddx}[2]{\frac{{\rm d} #1}{{\rm d} #2}}
\newcommand{\eq}[1]{{Eq.\,~(\ref{#1})}}
\newcommand{\respt}[1]{({\it resp.\ }#1)}
\newcommand{\eg}{{\it e.g.\ }}
\newcommand{\ie}{{\it i.e.\ }}

\newcommand{\Sect}[1]{{Sect.\,~\ref{#1}}}
\defcitealias{m97}{Paper~I}
\journalname{Astrophysics and Space Science (CoRoT/ESTA Volume)}

\begin{document}
\nocite{*}

\title{CESAM: a free code for stellar evolution calculations}

\author{P. Morel \and Y. Lebreton}

\institute{P. Morel\at Cassiop\'ee, URA CNRS 1362, Observatoire de la C\^ote 
d'Azur, Nice, France\\
Tel.: +033-492003039\\
Fax:  +033-492003121\\
\email{Pierre.Morel@obs-nice.fr}
\and
Y. Lebreton \at Observatoire de Paris, GEPI, CNRS UMR 8111, 5 Place Janssen, 92195 Meudon, France\\
\email{Yveline.Lebreton@obspm.fr} }

\date{Received date / accepted date}

\maketitle

\begin{abstract}
The \cesam code is a consistent
set of programs and routines which perform calculations of 1D quasi-hydrostatic
stellar evolution including microscopic diffusion of chemical 
species and diffusion of angular momentum.
The solution of the quasi-static equilibrium is performed
by a collocation method based on piecewise polynomials approximations projected
on a B-spline basis; that allows stable and 
robust calculations, and the exact restitution of the solution, not only
at grid points, even for the discontinuous variables. 
Other advantages are the monitoring by only one parameter
of the accuracy and its improvement by super-convergence.
An automatic mesh refinement has been designed for adjusting the localisations
of grid points according to the changes of unknowns.
For standard models, the evolution of the chemical composition is solved by 
stiffly stable schemes of orders up to four; in the convection zones
mixing and evolution of chemical are simultaneous. The 
solution of the diffusion equation employs the Galerkin finite elements scheme;
the mixing of chemicals is then performed by a strong turbulent
diffusion. A precise restoration of the atmosphere is allowed for.

\end{abstract}

\section{Introduction to CESAM.}
Within the limitations due to electronic degeneracy, \cesam allows
the computation of the quasi-static evolution of stellar models as long as the
assumption of quasi-static equilibrium remains valid, that is to say, 
until the exhaustion of oxygen in the core.
The modular structure of \cesam facilitates the choices among 
several physical formalisms, for equation of state (hereafter {\small EOS}), convection,
opacities, diffusion coefficients, \etc
Many nuclear networks and initial mixtures are available which allow
to optimise the physical description according to the kind of model
and evolutionary phase of interest.
Mass loss and infall of planetoids are also implemented.\\

\paragraph{The available packages.}
Earlier versions, {\sc cesam2-3-4}, were
programmed in F77 and {\sc cesam5} in F90. Though obsolete,
{\sc cesam4} and {\sc cesam5} are available at:

\centerline{\tt\small http://www.obs-nice.fr/morel/CESAM}
  
\noindent In the early 2000's, \cesam was re-programmed in F95 and named
 {\sc cesam2k}. Three versions are now available: 
 \begin{itemize}
 \item The ``{\em fixed version}'' and the ``{\em fixed COROT version}'',
 limited to 3$\alpha$ burning, available respectively at:
 
 \centerline{\tt\small http://www.obs-nice.fr/cesam/} 
 
 \centerline{\tt\small http://perso.obspm.fr/$\sim$lebreton/Modeles/CESAM.html}

 \item The ``{\em $\beta$ version}''  more complete, not fixed, still in
 development (not free of bugs) is hereafter signalled by the flag $(\beta)$. 
It includes evolution up to oxygen burning, diffusion of the angular 
momentum and other developments of minor importance. It is available at:
  
  \centerline{\tt\small http://www.obs-nice.fr/morel/CESAM}
  \end{itemize} 
 Each package contains five directories and two files:
 \begin{itemize}
  \item ``{\tt SOURCE}'', contains the {\sc fortran} sources.
  \item ``{\tt EXPLOIT}'', contains programs to exploit
  the models and examples of input files. 
  \item ``{\tt SUN\_STAR\_DATA}'', contains physical data and
  programs for their implementation.
  \item ``{\tt TESTS}'', contains programs performing various checks.
  \item ``{\tt SCRIPTS}'', contains scripts to be used
  for the implantation and operating, and a {\tt MAKEFILE}.
  \item ``{\tt aide\_mem2k.ps}'', a short guide of directions for use,
  the ``{\em aide-m\'emoire}'' (hereafter Paper 2).
  \item ``{\tt cesam2k.ps}'', a complete description of the numerical aspects and 
  physics implemented, the ``{\em notice}'' (hereafter Paper 3).
 \end{itemize}
The source is structured in 14 {\em modules}:
numerical routines, opacities, convection, etc. 
All the F95 routines of the source are compiled once. 
The requirements for the calculations are read in external files to be supplied
by the user. The run is interactive.
Messages displayed in French, or in English, allow the control of calculations.
The extent of convection zones,
a H--R diagram, and the profiles of temperature, pressure, luminosity and
abundances are displayed\footnote{Use of the {\tt PGPLOT} package.}
on line. \cesam has been especially designed to facilitate
the implementation of various physical constants,
opacities, {\small EOS}, atmosphere, nuclear networks etc.
So, its overall structure is separated in two spaces:
\begin{enumerate} 
\item A {\em``physical space''} where the coefficients of the differential
equations are written in a form close to their physical formalism.
\item A {\em ``numerical space''} where the differential equations are formally 
solved.
\end{enumerate}
Therefore \cesam allows to implement physical processes, and physical data,
without any knowledge of numerical methods involved for the solution of the
equations.
For the physics, the use of generic routines makes the reading of the
algorithms easier.

\paragraph{Units and values of physical constants.}
\cesam uses {\tt cgs} units except for the mass, radius and luminosity
expressed in solar units. Two sets of fundamental constants
are implemented which correspond to 
widely used values
\citep{c68,c88,li94,c20}.
For each calculation, one of these
sets is chosen as the unique source of fundamental constants.
Other constants are initialised locally, \eg the  
mass excesses, in the routine performing the calculations of
thermonuclear reaction rates.

\paragraph{Input files.} Very often only the 
 {\em ``input data file''} (hereafter IDF) is needed.
 It is read at the onset of the run and collects all the requirements needed for the calculations:
\begin{itemize}
 \item physical parameters: mass, chemical composition, mi\-xing-length
 parameter, etc.
 \item numerical parameters: maximum number of shells, kind of precision, etc.
 \item criteria for halting the computations: age to be reached, value of the
 hydrogen abundance at centre, etc.
 \item names and locations of the external
 data files containing the data of the tabulated {\small EOS} and opacity,
 names of the physical routines to be used, 
 name of the model, of the set of units to be used, etc.
\end{itemize}
The other input files
have only specific functions, among them the most useful are:
\begin{itemize}
 \item ``{\tt mixture}'' allows the use of an initial mixture not implemented
 in \cesamp
 \item``{\tt modif\_mix}'' allows to modify the abundances of some species
 in a mixture already implemented.
 \item ``{\tt reglages}'' allows to personalise the kind of precision to be used
 (see below).
 \item ``{\tt planet}'' contains the characteristics of infall of planetoids.
 \item ``{\tt rap\_iso}'' allows to modify the isotopic ratios. 
 \item ``{\tt vent}'' defines the chemical composition of the wind when it differs
 from the atmosphere composition.
 \item ``{\tt zoom}'' allows to fit the resolution of the display for the plot
  on line.
 \item ``{\tt langue}'' allows to have the messages in English.
\end{itemize}
The meaning of all items are explained in Paper~2.
Examples of files are given in the directory {\tt EXPLOIT}.

\paragraph{Output files.}
At the end of each time step, a {\em ``return binary file''}
(hereafter RBF) is created. It contains all the data needed to initialise
or pursue a computation.
There is the possibility, either to save all RBF, or to keep only
the last RBF created.
On request, {\em ``output data files''} (hereafter ODF) are created.
Three ODF are designed for adiabatic, non adiabatic
and inversion asteroseismic investigations. 
These ODF also serve as input for some programs of the directory {\tt EXPLOIT}. 
An ODF concerns the diffusion of the angular momentum ($\beta$).
All items of output  files are detailed in Paper~2.
It is also possible to create personalised ODF.

\paragraph{The kinds of precision.}\label{sec:preci}
To optimise the calculations, sets of parameters, named {\em ``r\'eglages''}, 
are fixed according to the kind of models to be calculated
and to their subsequent use. The most useful {\em r\'eglages} are:
\begin{itemize}
\item ``{\tt realistic precision}'' for standard evolutions.
\item ``{\tt super precision}'' used when a high level accuracy
is needed.
\item ``{\tt solar accuracy}'', close to {\tt super precision}, but especially
designed for seismological investigations; 
the number of shells of the last model is increased up to its maximal value.
\item ``{\tt corot}'', close to {\tt super precision}, but especially designed
for investigations connected to CoRot.
\item ``{\tt advanced}'', for the computation of early type star
models evolved up to the oxygen burning.
\item ``{\tt normal precision}'', for exploratory work.
\item ``{\tt low mass}'', for the computation of late type star models.
\item ``{\tt reglages}'', in that case, the parameters, designed by the user, are read on an input file named
``{\tt reglages}''.
\end{itemize}
The larger the expected accuracy, the larger the computational
expense.

\paragraph{Operating data and programs.}
The directory {\tt EXPLOIT} contains:
\begin{itemize}
\item examples of input files as IDF, ``{\tt reglages}'', ``{\tt planet}'',
``{\tt modif\_mix}'', etc.
\item ASCII files of preliminary models for the initialisations of PMS and ZAMS
models.
\item miscellaneous programs to make plots of the chemical composition profile,
or an extension of the grid on a given set of radius,
or create the IDF for solar calibration, etc.
\end{itemize}
The flow chart of \cesam is described in \Sect{sec:flow}.
As the numerical features are detailed in the appendix
of \citet{m97}\footnote{Only available on electronic form.} (hereafter Paper I)
they are only succinctly recalled in \Sect{sec:num}, except for
the automatic allocation of mesh points described in \Sect{sec:mvgrid}.
The restitution of the atmosphere is outlined in \Sect{sec:atm}.
The algorithms performing the temporal evolution are described in
\Sect{sec:evol}. The nuclear network is detailed in \Sect{sec:chem} and the
implementation of the rotation is described in \Sect{sec:rot}.
The various formalisms of convection implanted in \cesam are described in
\Sect{sec:conv}. Mass loss formalisms and infall of planetoids are described in 
\Sect{sec:loss}. {\small EOS} and opacities data available are listed in
\Sect{sec:eos}. 

\section{The flow chart.}\label{sec:flow}
\paragraph{Initialisations.}
At the onset of the run, the IDF is read.
Chemical composition is initialised
according to the initial mixture and to the isotopes used by the chosen nuclear network. 
Then, using fit-formulas (see \Sect{sec:nuc}), the thermonuclear reactions rates 
are tabulated on a relevant interval of
temperatures\footnote{The errors
introduced by these interpolations remain within the error bars of the data.}.
Then, evolution begins:
\begin{itemize}
\item {\bf either} it starts from zero age on {\small PMS} or {\small ZAMS}: 
an initial model having the required specifications
is deduced from a model taken from a RBF or from a model in {\small ASCII} chosen in
the directory {\tt EXPLOIT}. 
\item{\bf or} it pursues a previous calculation, then the input is one
RBF of the evolution going on. 
\end{itemize}

\paragraph{The evolution.}
The number of shells is updated as explained  in \Sect{sec:mvgrid}. Then,
taking possible overshooting into account, the limits between radiative zones
and convective mixed zones (hereafter LMR) are localised.
The angular velocity ($\beta$) and the chemical composition are then updated.
{\em In fine}, the equations of quasi-static equilibrium for the interior
and the atmosphere are solved. The process is repeated until convergence.

When the nuclear engine is at work, the time step control is first
based on the local accuracy achieved for the numerical integration 
of species of relevant interest (see \Sect{sec:nodiff} and \Sect{sec:diff}) and, second on a 
limitation of the relative changes of helium on the whole star.
Otherwise, during the pre-main sequence, the
limitation of the change of mechanical energy controls the time step.
In case of divergence of any iterative algorithm the time step is halved.

\paragraph{Stop criteria.}
According to flags, read in the IDF, the computations may be stopped:
\begin{itemize}
\item when the expected age is reached.
\item  when the central temperature reaches a given value.
\item  as soon as the abundance of hydrogen at centre reaches a given value. 
\item at the exhaustion of hydrogen at centre.
\item  when the helium core reaches a given extent.
\item  when the effective temperature crosses a given value.
\item at the ignition of the $3\alpha$ cycle.
\item  at the ignition of the carbon cycle.
\item  at the ignition of the oxygen cycle.
\end{itemize}
For most kinds of precision, the last time step is adjusted in order to 
fulfil the required stop condition. 

\section{Numerical methods.}\label{sec:num}
\paragraph{Choice of variables.}
For the numerical integration of the stellar structure equations,
the Lagrangian form is the most convenient as the discretisation
on mass is readily expressed.
However, it presents a singularity at centre and the core needs to be
integrated apart. The Eulerian form of the equations
does not suffer from such inconvenience,
but since the stellar radius varies with time, there is 
a free boundary \citep[see][par.~7.3]{sb79}.
With the Lagrangian variables:
$\mu\equiv M^{2/3}$, $R^2$, $L^{2/3}$, the central singularity disappears
\citepalias[see][par.~B1]{m97} and there is no need of a special
treatment for the core ($M$ is the mass, $L$ the luminosity and
$R$ the radius).
To be consistent, the chemical species are taken as functions of $M^{2/3}$.
The pressure, $P$, and the temperature, $T$, are expressed in logarithms,
($\xi=\ln P$, $\eta=\ln T$), on the ground that they change by more than six
magnitudes from the centre to the atmosphere.

\paragraph{Solving the differential equations.}
The unknowns are approached by piecewise polynomials of order defined 
according to the required accuracy; the mostly used are of order 1,
\ie linear piecewise, and of order 2, \ie parabolic piecewise.
For the stellar modelling, such a flexible representation is well adapted to the
presence of discontinuities resulting from the mixing of the convection zones.
For the calculations, the piecewise polynomials are projected on a local
linear basis of normalised B-splines \citep{db78,sk81}. That allows
to find back  exactly the solution at any location.
Moreover, B-splines basis are also used for solving:
\begin{itemize}
\item the two points boundary initial value problems of the stellar
structure and of the atmosphere by collocation  \citep[][ch. XV]{db78}.
\item the diffusion equations of chemicals
 and angular momentum by finite elements \citep{qv94}. 
\end{itemize}
The linear
systems involved by the resolution of implicit equations are band-diagonal.\\ 
However, due to the non-trivial and 
unfamiliar algebra of B-splines, the algorithms are much more
elaborated than with the finite differences. Furthermore, efficient and
stable algorithms have been constructed for integration, differentiation,
integration of differential equations by collocation
and for interpolation. In \cesam the routines,
especially constructed to manage the calculations with B-splines,
are derived from the algorithms of \citet[][chap. 4]{sk81}.
Details are given in \citetalias{m97}.

\subsection{Moving grid, mesh refinement and discontinuities
tracking.}\label{sec:mvgrid}
An automatic mesh refinement is implemented.
At time $t$, the locations of the $n(t)$ mesh points are set by
fulfilling the condition that, from a grid point to the 
next, the jump of a strictly monotonous {\em ``repartition function''},
$Q(\mu,t)$, is equal to a {\em ``repartition constant''} $C(t)$ \citep[see][sect. 16.5]{eg71,pf86}.
The locations of grid points, $\mu_i,i=1,\ldots ,n$, known at the issue of
the computations, satisfy:
\begin{equation}\label{eq:eqq}
Q(\mu_{i+1},t)-Q(\mu_i,t)\equiv C(t),\quad i=1,\ldots ,n-1.
\end{equation}
The choice of $Q(\mu,t)$ is based on an {\it a priori} knowledge of the
behaviour of the solution. For each $t$, one defines an 
``index'' function  $q(\mu,t)$ mapping 
$[0,\mu_{\rm b}]$ on $[1,n]$; the index 1 \respt{$n$}
corresponds to the centre and the index $n$ to the surface,
\ie $\mu_{\rm b}=M_{\rm ext}^\frac23$.
Therefore, the integration is made on an {\em equidistant grid}.
In terms of the derivative of $Q$ with respect to $q$, \eq{eq:eqq} reads:
\begin{displaymath}\label{eq:dpsi}
\left( \dsdx \psi q \right)_t=0,
\ {\rm with}\ \psi(t)\equiv\left(\dsdx Qq\right)_t.
\end{displaymath}
The change of variables $\mu\rightarrow q(\mu,t)$:
\begin{displaymath}
 \psi(t)=\theta\left(\dsdx\mu q\right)_t,
\ {\rm with}\ \theta(\mu,t)\equiv\left(\dsdx Q\mu\right)_t
\end{displaymath}
is calculated from the analytic form of $Q(\mu,t)$.
There are two more unknowns:
$\psi(t)$ and $\mu(q,t)$; they fulfil a system
of differential equations of first order with boundary conditions:
\begin{displaymath}\label{eq:muq}
\left(\dsdx \mu q \right)_t=\frac\psi \theta ,
\ \left(\dsdx \psi q \right)_t=0,\ {\rm with}\ 
\left\{
\begin{array}{l}
q=1,\ \mu=0 \\
q=n,\ \mu=\mu_{\rm b}.
\end{array}
\right.
\end{displaymath}
The differential equations of internal
structure and of atmosphere, written with respect to $q$, 
are detailed in \citetalias{m97}.
The equations are then solved on an {\em equidistant} grid, that allows
numerical optimisations.

\paragraph{Choice of $Q$.}
$Q$ should be a strictly monotonous, two times differentiable 
function as simple as possible. By experiments, 
it has been found that the most convenient compromise is:
\begin{displaymath}\label{eq:qrp}
 Q=\Delta_\xi\xi+\Delta_\eta\eta+\Delta_\mu\mu,
\ \Delta_\xi=\Delta_\eta= -1,\ \Delta_\mu= 15.
\end{displaymath}
where $\xi$ and $\eta$ have been defined above.
 
\paragraph{Mesh refinement.}
The initial value of $C(t)$ is fixed according to the
expected level of accuracy. Along the evolution, its value is kept within $\pm 
2\%$, of its initial value by increasing (or decreasing) 
the total number of shells. 

\paragraph{Setting a grid point on a LMR.} 
At the limit between a mixed zone and a radiative one,
there may be a singularity of chemicals. Therefore each LMR
has to coincide precisely with a grid point. To do that \cesam uses a weighted
repartition function. On each side of a LMR, the weights are computed in
such a way that they adjust locally the values of $C(t)$ to the amounts
just needed for iteratively ``pushing'' the closest grid point on the limit.
In most cases, including mixing, the distances from the LMR locations to the 
closest grid points are lower than a few per cents of the characteristic local grid size.
The more well defined the location of the LMR, the most efficient the
algorithm.

\paragraph{The grids.}
\cesam uses several grids for the B-splines representations of
quasi-static variables, atmosphere and  rotation variables. 
As it coincides with the {\em adjustable} grid of quasi-static variables,
the Lagrangian
grid used for the abundances of chemicals, is not fixed with respect to time.
There is the possibility to use a {\em ``fixed grid''}, to avoid the numerical
diffusion resulting from the variable Lagrangian grid.

\section{Atmosphere.}\label{sec:atm}
 The atmosphere connects the convective optically thick
outer part of the envelope to the optically thin interstellar medium.
In the interior, the diffusion approximation
\citep{kw91} is used to simplify
the calculation of the radiative flux. In the outermost parts, this
approximation is no longer valid, as soon as the Rosseland optical depth
is lower than $\simeq 10$: a special treatment is needed to restore the
atmosphere.\\
\cesam restores the atmosphere from a 
$T(\tau, T_{\rm eff}, g)$ law here, $\tau$ is the Rosseland optical depth,
$T_{\rm eff}$ the effective temperature and $g$ the gravity.
The stellar radius, $R_\star$, is defined as the bolometric one, {\em i.e.}
the radius at the
level where the local temperature is equal to $T_{\rm eff}$ \citep{mp4}. 
For genuine radiative $T(\tau)$ laws, whatever is $T_{\rm eff}$,
$\tau_\star$  is located at a fixed optical depth, {\em e.g.}
$\tau_\star=2/3$ for the Eddington's law. For more precise laws, the value of $\tau_\star$
is not fixed. Therefore, the location where $\tau_\star$ is
defined is a free boundary.\\
As oscillation modes are reflected in the outermost parts of stars,
the pressure, the temperature and their gradients should be continuous
at the limit with the envelope. The
continuity of the pressure gradient is trivially insured by the equation
of quasi-static equilibrium verified on both sides of the limit. 
The continuity of the temperature gradient is more intricate to fulfil
as the connection occurs in zones of convective instability.
For the restoration of the atmosphere, the temperature gradient is
derived from the
$T(\tau)$ law itself. Fixing the gravity the $T(\tau)$ law is expressed as:
$T^4=\frac 34 T_{\rm eff}^4f(\tau)$.
\[{\rm Using}\ d\tau=-\kappa \rho dr\ {\rm and}
\ \nabla_{\rm rad}=\frac3{16\pi acG}\frac{P\kappa L}{MT^4},\] after some
calculations:
\begin{eqnarray}\label{eq:grad}
&&\nabla\equiv\frac{d\ln T}{d\ln P}=
\frac{P\kappa}T\ddx T\tau\frac{R^2}{GM}=\\
&&\frac 3{16\pi ac G}\frac{P\kappa L_\star}{MT^4}
\left(\frac R{R_\star}\right)^2 \ddx f\tau=
\nabla_{\rm rad}\frac L{L_\star}\left(\frac R{R_\star}\right)^2\ddx f\tau
\nonumber
\end{eqnarray}
Here $\kappa$ is the Rosseland mean opacity, $\rho$ the density,
$a$ the radiation constant, $G$ the gravitational constant, $L_\star$ the
luminosity of the star, $c$ the speed of light and $\nabla_{\rm rad}$
the radiative gradient.
In principle, the continuity is ensured when the same convection theory
prevails on both sides. As in most cases the values differ,
the continuity of the temperature gradient is insured by
a weighted mean with respect to $\ln\tau$.\\
Eq.~(\ref{eq:grad}) is no longer valid for a genuine radiative $T(\tau)$
law, as it ignores convection. In such a case,  
following the prescription of ~\citet{hvb65} 
\citetext{M. Gabriel, J. Christensen-Dalsgaard, priv.\ comm.},
the temperature gradient in the convective atmosphere is computed with a
modified radiative gradient:
\[\nabla_{\rm rad}^*=\nabla_{\rm rad}\frac{df}{d\tau}.\]
For a genuine radiative $T(\tau)$ law:
\[\lim_{\tau \gg 1}\frac{df}{d\tau}=1,\] therefore,
at the limit between the atmosphere and the envelope, the radiative gradient
is continuous and consequently the convective temperature gradient.\\
The numerical integration of the differential equations fulfilled in the
atmosphere is made by collocation, see
\citetalias{m97} for details. The number of shells in the
atmosphere is fixed from 50 to 100 according to the required level of accuracy.

\section{Evolution of the internal structure.}\label{sec:evol}
\paragraph{Initial PMS model.}
The energy source in an initial PMS model is only of gravitational origin.
At the onset of the Hayashi track, the star is fully convective, therefore
isentropic and chemically homogeneous. The energy equation is reduced as in
\citet{i65}:
\begin{equation}\label{eq:pms}
\dsdx LM=\epsilon_G=-T\dsdx St=c(t)T
\end{equation}
where $c(t)$ is the ``contraction constant'', and $S$ the entropy.
Along the interval of time
$\Delta t$, the energy radiated equals the change of gravitational
energy, therefore, at first order:
\begin{eqnarray*}
&&\frac{L(t)+L(t+\Delta t)}2\Delta t\sim\left(\frac{GM^2}{R(t+\Delta t)}
-\frac{GM^2}{R(t)}\right)\ \Rightarrow\\
&&\Delta t\sim\frac{2GM^2\left[R(t)-R(t+\Delta t)\right]}{(L(t)+L(t+\Delta t))
R(t)R(t+\Delta t)}.
\end{eqnarray*}
here $R(t)$ is the stellar radius. 
An estimate of the initial time step is deduced from two models computed with
\eq{eq:pms}, with close values of the contraction
constant, $c(t)$ and $c(t+\Delta t)$.
\cesam uses: $c(t+\Delta t)=1.1\times c(t)$. With $c(t)=0.02$\ {\tt cgs} the
temperature at centre is about $10^5$K; it is ten times larger with
$c(t)=0.00008$\ {\tt cgs}.
Changes of the contraction constant allow to choose between different initial PMS
models. Most of the PMS models can be initialised with a preliminary
model in ASCII available in the directory {\tt EXPLOIT}.
\cesam assumes that a PMS model becomes a ZAMS model as soon as the release of
gravitational energy balances the thermonuclear nuclear one.
Such a model is chemically inhomogeneous.

\paragraph{Initial homogeneous ZAMS model.}
A model of ZAMS with homogeneous chemical composition is not a physical reality
as the nuclear engine does not work at equilibrium. However, it is a very
convenient short way as, after a few time steps, the model is very close
to the model at the end of the PMS. Several ZAMS initial models in ASCII are
available in the directory {\tt EXPLOIT}.
\section{Evolution of chemicals.}\label{sec:chem}

\subsection{The nuclear networks.}\label{sec:nuc}
At the onset of the computations, the abundances of chemicals are initialised
according to the initial hydrogen and helium mass ratios and mixture.
The initial abundance of each chemical species is split between its isotopes,
according to the isotopic ratios of nuclides.
Several mixtures are implemented, the most useful are: the 
 \citet{ag89} meteoritic mixture and the solar mixtures of 
 \citet{gn93} and \citet{gs98}.
If necessary an IDF allows to modify the initial abundances of specific
species or to use a mixture not yet implemented. \\
Up to 16 nuclear networks are presently available. Hence, one can follow 
the evolution using only the chemical species and the thermonuclear
reactions of interest.\\
The nuclear reaction rates are tabulated
on relevant intervals of temperatures. The rates are computed using the
formulas of \citet{cf88} or of NACRE compilations \citep{a99}. For solar models, the 
improved rates of \citet{al98} are available.
The weak screening of \citet{s61} and the weak  and
intermediate screenings of \citet{mi77} are available. \\

\subsection{Evolution without diffusion.}\label{sec:nodiff}
The time scales involved in the temporal evolution
of chemicals differ by a large number of magnitudes.
From a mathematical point of view it is a {\em stiff} problem \citep{hw91} and algorithms have been especially designed for it. 
Without microscopic diffusion, L-stable
\citep[][par. 4.3]{hw91} implicit Runge-Kutta schemes,
are available for the chemical evolution.\\
In the radiative zones, the equations to be solved are formally written:
\begin{equation}\label{eq:chim}
 \dsdx{x_{\rm i}}t =  \Psi_i(P,T,{\cal X},t),\ 1\leq i\leq n_{\rm X}.
\end{equation}
Here, $x_{\rm i}$ is the abundance per mole of the chemical species 
$i$, $\Psi_i$ the rate of change of $x_i$, ${\cal X}$ the vector
of chemical abundances, $t$ the time and $n_{\rm X}$ the number of chemicals;
$x_{\rm i}$ is given by:
\[x_{\rm i}=\frac{X_{\rm i}}{\nu_{\rm i}},\]
$X_{\rm i}$ is the abundance mass fraction and 
$\nu_{\rm i}$ the atomic mass. In mixed zones (hereafter MZ), convective eddies homogenise the chemical composition.
There, the mixing and updating of
chemicals are done simultaneously, therefore the changes of mean
abundances read:
\begin{equation}\label{eq:dx2}
 \ddx {\bar{x_i}}t=\int_{\rm MZ}\Psi_i(P,T,\bar{\cal X},t)
 {\rm d} M \bigg/  \int_{\rm MZ} {\rm d} M. 
\end{equation}
As a grid point is defined on each LMR (see \Sect{sec:mvgrid}),
the discontinuities of the abundances are explicitly calculated. 

\paragraph{Control of the accuracy.}
A good estimate of the numerical accuracy of an integration is obtained
with the Fehlberg method \citep[][par. 7.5.2]{sb79}. It needs to triple the
calculations. As it is prohibitive, falling anything better, the time
step is simply adjusted in such a way that, over a time step, the relative
changes of the abundances remain within fixed limits. The largest the expected
accuracy, the narrower the limits..

\subsection{Evolution with diffusion.}\label{sec:diff}
\label{sec:mixd}
With microscopic diffusion, the equations of the evolution of chemicals
have the form:
\begin{eqnarray}\label{eq:diffu}
\frac{\partial x_i}{\partial t}&=&\frac{\partial F_i}{\partial M}+
\left(\frac{\partial x_i}{\partial t}\right)_{\rm nucl.}\\
F_i&=&4\pi R^2\rho\left(4\pi R^2\rho D_i\bullet\nabla\!_{\rm M}
{\cal X}+v_i x_i\right)\nonumber
\end{eqnarray}
here, $\nabla$ is the gradient operator and $v_i$ the advection velocity. 
The symbol ``$\bullet$'' means the vector inner product 
\[{\mathrm{and}\ }\nabla\!_{\rm M}{\cal X}=\left(\dsdx{x_1}M,\ldots,
\dsdx{x_{n_X}}M\right)^{\rm T}.\]
The turbulent diffusion coefficients, are added to the $i$-th component of
the vector, $D_i$, of diffusion coefficients of the species $i$, see \eq{eq:di}.\\
For the integration, the abundances are approached
by piecewise polynomials expressed on a
B-spline basis with discontinuous derivatives at each LMR.
A finite-elements method \citep[see \eg][]{qv94}
is used to solve the diffusion equation. 
That allows an integration by parts which reduces to unity
the order of the diffusion equation. The scheme is fully implicit.
The nuclear term is evaluated as for the implicit Euler's formula.\\
The mixing is made by turbulent diffusion with coefficient $D_{\rm MZ}>>1$.
At each LMR, the abundances $x_i$ and fluxes $F_i$ are continuous functions 
with discontinuous first derivatives, owing to the jumps of the diffusion
coefficients. Therefore \eq{eq:diffu} holds everywhere.
Two formalisms are available for the calculation of the diffusion vector $D_i$:
\begin{itemize}
\item The coefficients are calculated according to \citet{mp93}.
The metals are ``test elements'', their diffusion
only results from collisions against protons. Based on the presence of protons,
this formalism is only valid for the main sequence.
\item The diffusion coefficients are computed according to
\citet{b69}, this formalism is outlined beneath.
\end{itemize}

\paragraph{Boundary conditions.}
At the outermost limit, $M=M_{\rm ext}$, it is assumed that there
 is neither input nor output of matter, then $F_i(M_{\rm ext})=0$
for any particle $i$. At centre $M=0$, because of the spherical symmetry,
$F_i(0)=0$. 

\paragraph{Control of the gravitational settling.}
For stellar models with mass larger than $\approx 1.4\,M_\odot$,
the use of microscopic diffusion {\em alone} produces
an important depletion of helium and metals at the surface  and a concomitant enhancement of hydrogen.
Different ways to overcome this problem have been used. 
\citet{ecb05} introduce some turbulence due to rotation. \citet{dm04}
suppress diffusion in the outer layers. \citet{cdg99} include a
wind mass loss which reduces the diffusion in outer layers,   
\citet{trm98} introduce a turbulent mixing. \cesam allows to control gravitational settling with a
radiative turbulence of coefficient $d_\nu$ 
proportional to the radiative kinetic viscosity; it results from
the energy exchanges between thermal collisions leading to excitation and
ionisation of atoms and ions \citep[][p. 461-472]{t30,mm84}.
\begin{equation}\label{eq:renu}
d_\nu=Re_\nu\frac4{15}\frac{aT^4}{c\kappa\rho^2},
\end{equation} 
$c$ is the speed of light in vacuum.
The phenomenological parameter $Re_\nu$ has been found close to unity by
\citet{mt02}. The physical meaning of this, as efficient as simple source
of turbulent mixing, has been questioned by \citet{am05}.

\paragraph{Burgers's flow equations.}\label{sec:bg}
With respect to the abscissa $x$,
the density number $n_i$ and $w_i$, the diffusion velocity of the particle $i$,
are related by \citet{id85}:
\[\dsdx{n_i}t=\frac 1{x^2}\dsdx{ }x(x^2n_iw_i)
+\left(\dsdx{n_i}t\right)_{\rm nucl.},\]
With the formalism of \citet{b69}, the diffusion velocity is expressed as:
\[w_i=\sum_j b_{ij}\dsdx{x_j}x+v_i.\]
where the diffusion coefficient
of the particle $i$ with respect to the particle $j$, $b_{ij}$,
and the advection velocity, $v_i$, come from the
solution of a linear system. The diffusion velocities, for ions and electrons,
\citep{b69,cgk89,tbl94},
satisfy\footnote{Without conduction current and magnetic field.}:
\begin{eqnarray}\label{eq:bu11}
\frac{dP_i }{d R}-\frac{\rho_i}\rho\frac{d P}{d R} &-& n_i\bar{z_i} eE
= \sum_j K_{ij}(w_j-w_i) + \nonumber \\
 &+& \sum_j K_{ij} z_{ij}\frac{m_jr_i-m_ir_j}{m_i+m_j} \\
\frac 52 n_i k \frac{d T}{d R} = &-& \frac 52 \sum_j K_{ij} z_{ij}
\frac{m_j(w_j-w_i)}{m_i+m_j}-\frac 25 K_{ii}z_{ii}^{\prime\prime}r_i
\nonumber \\
&-&r_i\sum_{j\not= i}K_{ij}\frac{3m_i^2+m_j^2z_{ij}^\prime\label{eq:bu22}
+\frac45m_im_jz_{ij}^{\prime\prime}}{(m_i+m_j)^2}\nonumber \\
&+&\sum_{j\not= i}K_{ij}\frac{m_im_j(3+z_{ij}^\prime
-\frac45z_{ij}^{\prime\prime})}{(m_i+m_j)^2} r_j.
\end{eqnarray}
$e$ is the electron charge, $E$ the electric field,
$P_i$ the partial pressure, $m_i=\nu_im_u$ is the mass
of particle $i$, $k$ the Boltzmann constant,
 and $r_i$ the magnitude of the residual heat flow vector;
$\rho_i$ the partial density is given by:
\begin{equation}\label{eq:roi}
\rho_i=\rho X_i=\rho x_i\nu_i=n_i\nu_im_u,
\end{equation}
with $N_0$ as the Avogadro number, the inverse of the atomic mass unit $m_u$:
\begin{equation}\label{eq:ni}
n_i=\rho N_0\frac{X_i}{\nu_i}=\rho N_0 x_i.
\end{equation}
The force equation~(\ref{eq:bu11}) represents the pressure and the concentration
dependence of the diffusion velocity, while Eq. (\ref{eq:bu22}) prevails
 for its thermal dependence.
The charge of any isotope is taken as the
averaged\footnote{Weighted by the ionisation rates.}
charge, $\bar{z_i}$, over all its ionisation states. An
unique mean charge is used for all the isotopes of a given chemical.
The quantities $K_{ij}$ are the so-called resistance
coefficients, they represent the effects of collisions between particles
 $i$ and $j$ \citep{mp93}:
\begin{equation}\label{eq:kij}
K_{ij}=\frac{16}3 n_i n_j m_{ij}\Omega^{(11)}_{ij},
\end{equation}
with $m_{ij}=(m_im_j)/(m_i+m_j)$ as the reduced mass of particles $i$ and $j$.
The heat flux terms involve additional collision integrals:
\[z_{ij}=1-\frac25\frac{\Omega^{(12)}_{ij}}{\Omega^{(11)}_{ij}},\]
\[ z_{ij}^\prime=2.5-\frac25\frac{5\Omega^{(12)}_{ij}-\Omega^{(13)}_{ij}}
{\Omega^{(11)}_{ij}},
\ z_{ij}^{\prime\prime}=\frac{\Omega^{(22)}_{ij}}{\Omega^{(11)}_{ij}},\]
\citet{ppfm86} showed that
$\Omega^{(kl)}_{ij}$ can be written:
\[\Omega^{(kl)}_{ij}\equiv F^{(kl)}_{ij}\epsilon_{ij},
\ (kl)=(11),\ (12),\ (13),\ (22),\]
\[ \epsilon_{ij}\equiv\frac{e^4}4\sqrt{\frac\pi{2k^3}}
\ \frac{\bar{z_i}^2\bar{z_j}^2}
{\sqrt{m_{ij}}}\ \frac1{\sqrt{T^3}}.\]
For attractive and repulsive screened Coulomb potentials,
the quantities $\ln(F^{(kl)})_{ij}$ have been tabulated by
\citet{ppfm86}. The equations of dynamical conservation of the mass and charges 
respectively are:
\begin{equation}\label{eq:cons}
\sum_{i=1}^{n_X}x_i\nu_iw_i=0,\ \sum_{i=1}^{n_X+1}\bar{z_i}x_iw_i=0,
\end{equation}
where $n_X+1$ is the index of electrons. 
Following \citet{ppfm86} and other works, \eg
\citet{id85}, \citet{cgk89}, \citet{tbl94}, the diffusion
velocities $w_i$ come from the solution of the  system
of $2n_X+3$ linear equations, formed by the
$2n_X$ \eq{eq:bu11} and \eq{eq:bu22} for the ions, the
\eq{eq:bu22} for the electrons and the two \eq{eq:cons}. The $2n_X+3$
unknowns are
$w_i,\ i=1,\ldots,n_X+1$, $r_i,\ i=1,\ldots,n_X+1$ and $E$.\\
For want of something better, with the ideal gas law, 
the pressure and the partial pressures respectively are:
\begin{equation}\label{eq:gp}
P=\frac{\rho{\cal R}T}\mu,\ P_i=n_ikT=\rho x_i N_0kT=P\mu x_i,
\end{equation}
here $\cal R$ is the perfect gas constant, and $\mu$ the mean molecular weight.
In spherical symmetry, the pressure and the temperature gradients are given by:
\begin{equation}\label{eq:dp}
\frac{d P}{d R}=-\rho g,\ \frac{d T}{d R} =
\frac T{\cal P}\frac{\partial \ln T}{\partial \ln {\cal P}}
\frac{d {\cal P}}{d R}=-\frac TP\nabla\rho g.
\end{equation}
Working with \eq{eq:roi} to \eq{eq:dp}, with respect to $\mu$ and $x_i$,
the Burgers's equations (\ref{eq:bu11}), (\ref{eq:bu22}), may be rewritten:
\[A\omega=\gamma+G\,D_x,
\ D_x=(\dsdx{x_1}R,\ldots,\dsdx{x_{n_X}}R,0,\ldots,0)^T.\]
The solutions are:
\[\omega=V+B\,D_x,\ V=A^{-1}\gamma,\ B=A^{-1}G.\]
For abridgment, neither the derivation
of above equations, nor the complicated forms of vector $\gamma$ and matrix
$A$ and $G$, are reproduced, all details are given in Paper~3.
The diffusion velocities of the ions are expressed as:
\[w_i=\sum_{j=1}^{n_X}b_{ij}\dsdx{x_j}R+v_i,\ i=1,\ldots,n_X,\]
and, owing to Eq. (\ref{eq:ni}), the diffusion vector reads:
\begin{equation}\label{eq:di}
D_i=\left(x_ib_{i1},\ldots,x_ib_{in_X}\right)^{\rm T},
\end{equation}
for $i,j=1,\ldots,n_X$, $b_{ij}$ and $v_i$, are respectively the coefficients
of matrix $B$ and vector $V$.

\paragraph{Calculation of mean charges.}\label{sec:saha}
The Burgers formalism involves the charges of the isotopes.
To simplify, \cesam considers a unique mean charge for all the isotopes
of each chemical. 
For the calculation of the ionisation rates, the Saha-Boltzmann equation
\citep[][eq.15-30]{cg68} has been adapted in the following way.
Let $n_{j,i}$ be the number density of atoms in ionisation state $j$ of
the chemical species $i$. The ratio of the total number of atoms in
successive stages of ionisation can be written:
\begin{equation}\label{eq:saha0}
\frac{n_{j-1,i}}{n_{j,i}}=\frac{U_{j-1,i}}{U_{j,i}}\exp\left(\eta
+\frac{\chi_{j,i}}{kT}\right),
\end{equation}
here $\chi_{j,i}$ is the ionisation potential, $U_{j,i}$ the partition
function and $\eta$ the electron degeneracy, related to the number density of
free electrons and temperature,
through the half integer Fermi-Dirac function \citep[][eq. 2-57]{c68}.
\citet{eff73} have introduced a convenient 
approximate treatment of pressure ionisation: a numerical correction is postulated to ensure that the plasma
remains completely ionised at sufficiently high density.
In the inner solar radiative zone, this approximate treatment
leads to too large
mean charges for the ions and, at the centre, 
iron is fully ionised, while it is only 85\% according to Table~1 in
\citet{g97}. A similar behaviour is observed with the
modified parameters recommended by \citet[][eq. 4]{pm91}.\\
In \cesamv the partition functions are limited to the
statistical weights of fundamental levels and, as soon as the mean
distance between the ions becomes of the order of the size
of the ion cloud, $R_{\rm D}$, the ratio of statistical weights is
smoothly reduced to zero.
The Debye-Huckel radius writes \citep[][eq. 2-235]{c68}~*:
\[R_{\rm D}=\sqrt{\frac{kT}{4\pi e^2\rho
N_0\zeta}},\quad\zeta=\sum_{i\not=e}\bar{z_i}(\bar{z_i}+1)x_i,\]
The Saha-Boltzmann Eq. (\ref{eq:saha0}) is then written: 
\begin{equation}\label{eq:saha1}
\frac{n_{j-1,i}}{n_{j,i}}=\frac{g_{j-1,i}}{g_{j,i}}f(x)\exp\left(\eta
+\frac{\chi_{j,i}}{kT}\right),
\end{equation}
\[x=\frac{\chi_{i,j}}{\chi^\prime_{j,i}}-1,
\quad\chi^\prime_{j,i}=\max(0,\chi_{i,j}-\chi_{i,j}^{\rm C}),
\quad\chi_{i,j}^{\rm C}=\frac{je^2}{R_{\rm D}},\]
the $g_{j,i}$ are the statistical weights of fundamentals.
For the smoothing function $f(x)$, \cesam uses the piecewise cubic
polynomial with zero derivatives at $x=0$ and $x=p$ ($p=4$):
\[f(x)=\left\{ \begin{array}{ll}
0&\mbox{if $x\leq 0,$} \\
\left(\frac xp\right)^2
\left[-2\left(\frac xp\right)+3\right]&\mbox{if $x\in[0,p],$} \\
1&\mbox{if $x\geq p,$}
\end{array}\right.\]
Then, as soon as $\chi_{i,j}\leq \chi_{i,j}^{\rm C}$, the quantity
$g_{j-1,i}/g_{j,i}f(x)$ goes to zero and the level $j-1$ becomes fully ionised.
The set of Saha-Boltzmann equations for all ions
is written as in \citet[][eq. 5-17]{m78}. They are solved by iterations
using a second order Newton-Raphson scheme.
\begin{table}
\caption[]{Comparison between the mean charges from the approximate solution 
and the precise equation of state of \citet{g97} (label ``G'') 
beneath the convection zone and at centre, for a calibrated solar
model, .
}\label{tab:barZ}
\begin{tabular}{cccccccccccccc}  \hline \\
&\multicolumn{2}{c}{T=2.2\,MK} & \multicolumn{2}{c}{centre}\\ \\
elements &$\bar Z_{\rm G}$&$\bar Z$&$\bar Z_{\rm G}$ &$\bar Z$ \\ \\ \hline \\
$\rm C$      & 5.89           &5.92    &  6.00           & 6.00    \\
$\rm N$      & 6.75           &6.82    & 7.00            & 7.00    \\
$\rm O$      & 7.47           &7.56    & 8.00            & 8.00    \\
$\rm Fe$     & 17.0           &16.7    & 24.2            & 24.2    \\
\\ \hline
\end{tabular}
\end{table} 
For the mean charges, quantities of main interest
in the present investigation, Table \ref{tab:barZ} reveals agreements better
than $\pm 3\%$ with the data of Table 1 of \citet{g97}.

\section{Rotation.}\label{sec:rot}
Rotation is considered in \cesam under the assumption of spherical symmetry. With non zero angular velocity, the
mean centrifugal acceleration affects the local gravity.
In the initial model, rigid rotation is assumed.
The initial angular velocity can be read from the IDF in different units:
\begin{itemize}
 \item radian/s.
 \item km/s, it corresponds to the rotational velocity of the outer
 part of the star. As the outer radius depends on the outer gravity,
 the initial model has to be iteratively adjusted to have the required
 outer velocity.
 \item days, that corresponds to the rotation period.
\end{itemize}

\subsection{Rotation without diffusion of angular momentum.}
Several options are available:
\begin{itemize}
 \item no rotation: the initial angular velocity must also be zero.
 \item solid-body rotation where the angular velocity is kept to
 its initial value. 
 \item rotation is solid and angular
 momentum is globally conserved. The angular velocity changes with respect to
 time, according to structural changes.	 
 \item  angular velocity changes
 according to the conservation of the local angular momentum. 
 The rotation is not solid, but in the mixed zones.
\end{itemize}
	 
\subsection{Rotation with diffusion of angular momentum.}
The two formalisms of \citet{tz97} and \citet{mz04} of the diffusion of angular
momentum ($\beta$) are implemented in
\cesamp The diffusion  coefficients of the angular momentum are computed
according either to \citet{p03} or to \citet{mpz04}.

\section{Mass loss and infall of planetoids.}\label{sec:loss}
Several formalisms of mass loss are implemented. 
The mass loss rate is either negative or positive, \ie increase of mass.
With diffusion, the chemical composition of the
input \respt{output} can differ from those of the outer convective zone,
from where the output \respt{input} is assumed to come from \respt{vanish}.
The mass loss rate, in unit of $M_\odot. {\mathrm yr^{-1}}$ is read in the IDF.  
The following options are implemented:
\begin{itemize}
\item standard mass loss.
\item solar mass loss, the mass loss is halted as soon as the mass
of the model reaches the solar value.
\item mass changes due to nuclear energy generation.
\item infall of planetoids ($\beta$). The characteristics
of the infall, duration, amount of terrestrial masses, chemical
composition of the planetoids, are read in a IDF.
\end{itemize}
Angular momentum losses are implemented but in validation ($\beta$).

\begin{table*}[htb!]
\caption{The sets of parameters used for the calibration of solar
models with CESAM2k. We used (i) two solar heavy elements mixtures,
the GN93 \citep{gn93} or the AGS05 \citep{ags05} mixture; (ii) two formalisms
for convection treatment, the BV \citep{b58} and CGM \citep{cgm96}
formalism, (iii) two $T(\tau)$ laws for the atmosphere, the 
Eddington grey law and $T(\tau)$ laws derived from ATLAS9 model
atmospheres \citep{kur92} calculated with the same convection formalism 
as the interior models \citep[see][]{skglv06} and (iv) two formalisms 
for the calculation of the microscopic diffusion of the elements, the
MP93 \citep{mp93} and B69 \citep{b69} formalism.}
\centering
\label{tab:sun-models}       
\begin{tabular}{ccccc}
\hline\noalign{\smallskip}
{\bf Model} &{\bf Mixture} &{\bf Convection}& {$\mathbf T(\tau)$} law & {\bf Diffusion} \\[3pt]
\tableheadseprule\noalign{\smallskip}
\bfseries{A}& {GN93} & {BV}&{ATLAS9} & {MP93} \\
\bfseries{B}& {GN93} & {BV}& {EDDINGTON} & {MP93} \\
\bfseries{C}& {GN93} & {BV}& {ATLAS9} &{B69} \\
\bfseries{D}& {GN93}  & {CGM}& {ATLAS9}& {MP93}\\
\bfseries{E}& {AGS05} & {BV} & {EDDINGTON} & {MP93}\\
\noalign{\smallskip}\hline
\end{tabular}
\end{table*}
\begin{table*}[ht!]
\caption{Properties of the five {\small CESAM2k} calibrated solar models. $Y_0$, $Z_0$ 
and $(Z/X)_0$ are the initial values of the helium and heavy elements mass
fractions and $(Z/X)$ ratio. $Y_{\rm e}$, $Z_{\rm e}$ are the present values
in the convective envelope while $R_{\rm e}$ is the radius at the basis of
the convective envelope in units of the solar radius. $\rho_{\rm c}$,
$T_{\rm c}$ and $X_{\rm c}$ are the central values of the density in 
$\rm g.cm^{-3}$, temperature in $10^6$\ K and hydrogen abundance. 
$\alpha_{\rm conv,int}$ is the value of the convection parameter 
in the interior ($\alpha=l/H_p$ where $l$ is the mixing length and $H_p$ 
is the pressure scale-height)}
\centering
\label{tab:sun-results}       
\begin{tabular}{ccccccccccc}
\hline\noalign{\smallskip}
{\bf Model} &$Y_0$ & $Z_0$ & $(Z/X)_0$& $Y_{\rm e}$ & $Z_{\rm e}$&$R_{\rm e}$&$\rho_{\rm c}$& $T_{\rm c}$&$X_{\rm c}$& $\alpha_{\rm conv,int}$  \\[3pt]
\tableheadseprule\noalign{\smallskip}
\bfseries{A}&$0.2735$&$0.0196$& $0.0278$&$0.2447$&$0.0181$&$0.7143$&$153.1$&$15.72$&$0.3387$&$2.42$\\

\bfseries{B}&$0.2735$&$0.0196$& $0.0278$&$0.2447$&$0.0181$&$0.7145$&$153.1$&$15.72$&$0.3387$&$1.76$\\

\bfseries{C}&$0.2741$&$0.0198$& $0.0280$&$0.2460$&$0.0180$&$0.7152$&$153.0$&$15.72$&$0.3388$&$2.40$\\

\bfseries{D}&$0.2735$&$0.0196$& $0.0278$&$0.2447$&$0.0181$&$0.7143$&$153.1$&$15.72$&$0.3387$&$0.77$\\

\bfseries{E}&$0.2637$&$0.0141$& $0.0196$&$0.2339$&$0.0129$&$0.7297$&$151.0$&$15.52$&$0.3578$&$1.72$\\

\noalign{\smallskip}\hline
\end{tabular}
\end{table*}

\section{Convection.}\label{sec:conv}
Two formalisms for the computation of the temperature gradient in the
convection zones are available:
the standard \citet{b58} mixing-length (MLT) formalism is considered with the
optical thickness of the convective bubble, and 
the \citet{cm91} formalism. The mixing-length parameter is read in the IDF.\\
Overshooting beneath and/or above the convection zones can be accounted for. The
overshooting parameters, scaled by the local pressure scale height, are 
read in the IDF. In overshoot regions, the temperature
gradient is set, either to the adiabatic or to the radiative gradient.
The convective zones and their extents by overshooting are homogenised,
see \Sect{sec:nodiff}.\\ 
Up to now specialised treatment of the semi-convection is not implemented
in \cesamp With microscopic diffusion,
in areas swept across by the backward movement of the border of a convective
core, the discontinuities of chemicals are assumed to be eroded only by 
diffusion. Without diffusion ($\beta$), the abundances profiles are
spatially linearly interpolated between
their values on the convective core and at the former location of the LMR.
That avoids a noisy behaviour of chemical gradients and, consequently,
of the profile of the \bvp

\section{Equation of state and opacity.}\label{sec:eos}
Four analytical {\small EOS} are implemented.
The most useful are \textsc{eff} \citep{eff73} and
\textsc{ceff} \citep{cd92}. 
Numerical {\small EOS} are available. The \textsc{mhd} tables \citep{d88}
and the \textsc{opal} 1993 and \textsc{opal} 2001 tables are used with
the {\small OPAL} interpolation scheme for tables with $Z = 0.01$ and $0.02$.\\
{\small OPAL} 1996 opacity tables are implemented, with Kurucz low-temperature values. 
The ratios between the abundances of heavy-elements are fixed at their initial
values, regardless of changes due to nuclear reactions and diffusion.
All {\small OPAL} tables of type 2, available today,
are implemented altogether with \textsc{z14xcotrin21}, the
package of A.I. Boothroyd.\\
The neutrinos emitted in nuclear reactions are taken into account,
it is assumed that they can freely escape the star.
Other physical processes related to neutrino such as Urca process,
plasma neutrino, pair neutrino \citep[][par. 18.6]{kw91} have only been
implemented in private codes derived from \cesamp\\
For temperature values above $T \geq 7\,10^7$K $\simeq 7$Kev, the plasma is 
fully ionised, so the Rosseland mean opacity is reduced to the
Compton scattering by free electrons \citep[][par. 16.6]{cg68}. 

\section{Calibration of the Solar model.}\label{sec:sun}

\begin{figure}
\centering
\resizebox*{\hsize}{!}{\includegraphics*{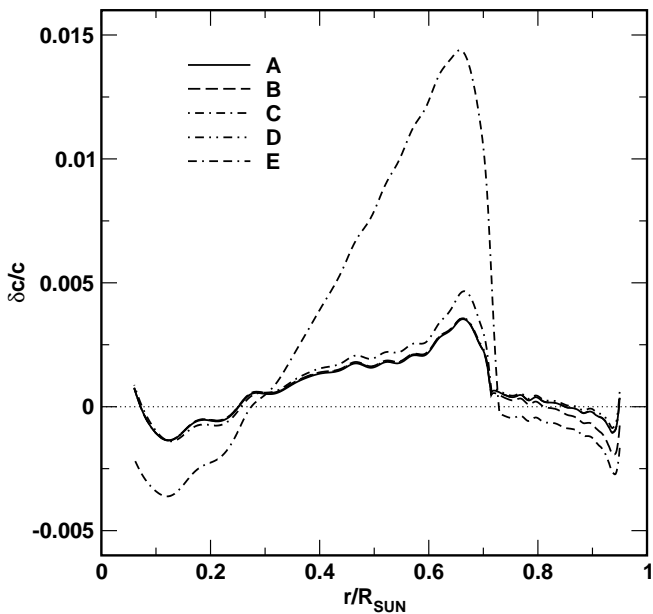}}
\caption{Relative differences between the seismic sound speed derived by \citet{basu00} and the solar models presented in Table~\ref{tab:sun-models}.}
\label{fig:sound}       
\end{figure}

We have performed calibrations of the solar model with {\sc cesam2k}
with various sets of input physics and initial parameters. This consists in adjusting the initial parameters of the model
(helium abundance and mixing-length parameter) in order to satisfy
the observational constraints on the solar global parameters at solar age. 

We have adopted the values of the astronomical and physical constants
specified for the calculation of the stellar models compared in the
different {\small ESTA} tasks \cite[see][]{yl-apss07}.  For the solar
global parameters, we therefore took
$R_\odot = 6.9599\  10^{10}\ {\rm cm}$ (radius),
$L_\odot=3.846\ 10^{33}\  {\rm{erg.s^{-1}}}$ (luminosity) and 
$M_\odot=1.98919\ 10^{33}\ {\rm g}$ (mass). The value $R_\odot$
refers to the radius of the model layer where $T=T_{\rm eff}=5777$\ K.
For solar age, we adopted the value
$t_\odot=4.57\ {\rm Gyr}$  \citep{bpw95}.

Table \ref{tab:sun-models} presents the sets of input physics
used in the different calibrations of solar models. All models have been
calculated with the {\small OPAL} 2001 {\small EOS}  \citep{rn02} and the
1995 {\small OPAL} interior opacities \citep{ir96}. Two sets of low temperature
opacities have been used: the \cite{af94} tables given for the \cite{gn93}
solar mixture and the \cite{ferguson05} tables given for
the new solar mixture derived by \cite{ags05} (see below). All models
take into account the diffusion of chemicals due to
pressure, temperature and concentration gradients (no radiative
accelerations) but we considered either the
\citet[][hereafter MP93]{mp93} or the \citet[][hereafter B69]{b69}
formalism. Convection is treated either according to the classical
{\small MLT} \citep[][hereafter BV]{b58} or to the
\citet[][hereafter CGM]{cgm96} formalism. In the CGM formalism,
like in the CM formalism \citep{cm91}, the contribution 
of eddies with different sizes is taken into account in the
calculation of the convective flux and velocity. In addition
\citet{cgm96} take into account the feedback of the turbulence
on the energy input from the source which generates turbulent
convection. For the atmosphere calculation we considered
either the classical Eddington grey $T(\tau)$-law or a
law derived from Kurucz's ATLAS 9 1-D model
atmospheres \citep{kur92}. We have taken   the same $T(\tau)$-laws
as used in the work by \citet{skglv06}. These laws are based on model
atmospheres calculated with either the BV or the CGM convection formulation.
In both cases, the atmosphere calculation was performed adopting a value of
the mixing-length parameter $\alpha_{\rm conv, atm}=0.5$ which allows to fit
at best the observed profiles of the solar Balmer lines \citep[see][]{vm96}.
Therefore $\alpha_{\rm conv, atm}$ is different from the value of
$\alpha_{\rm conv, int}$ in the interior, this latter being adjusted to
calibrate the solar model. Finally, we adopted
the GN93 solar mixture of heavy elements \citep{gn93} in all models but one
where we used the new AGS05 mixture \citep{ags05} which is derived from
a time-dependent, 3-D hydrodynamical model of the solar atmosphere. The
abundances of C, N, O of the AGS05 mixture are smaller than in the GN93
one which leads to an important decrease of the solar $(Z/X)$
ratio: $(Z/X)_{\odot}=0.0245$ for the GN93 mixture and $(Z/X)_{\odot}=0.0171$
for the AGS05 mixture.
 
Table \ref{tab:sun-results} presents the results of the solar model
calibrations. The relative differences in radius, luminosity and present
$(Z/X)$ surface value of the five models A, B, C, D, E with the observed
values are lower than $10^{-4}$. The relative differences between the seismic sound speed derived by \citet{basu00} and the models are plotted in Fig.~\ref{fig:sound}.
The helioseismically measured values of the present radius at the base of the
convective envelope $R_{\rm e,o}$ and of the present solar envelope helium
abundance $Y_{e,o}$ provide strong constraints for the solar model. \cite{ba97}
helioseismically derived $R_{\rm e,o}=0.713\pm0.001 R_\odot$.
\cite{bs03} derived a mean value  
$Y_{\rm e,o}=0.245\pm0.005$ from different helioseismic determinations.
In all our models but one (model E), we find values of $Y_{e}$ and
$R_{\rm e}$ in reasonable agreement with the seismic values. Model E is
based on the AGS05 solar mixture which makes the agreement between the solar
model and helioseismic observations much worse
\citep[see for instance][]{basu04}. More details on the solar models calculated with \cesam 
and their seismic properties can be found in \cite{m99,jp00,az07}.

\begin{acknowledgements}
We wish to express our thanks to A. Baglin and E. Schatzman for their
suggestion to 
undertake the development of \cesam and for their unfailing support.
Many contributions to \cesamv help, advice, debugging and constructive
criticisms are from our nearest colleagues:
G. Berthomieu, S. Brun, Th. Corbard, M.J. Goupil, A. Moya,
B. Pichon, J. Provost, F. Th\'evenin, C. van't Veer, J.P. Zahn.
We~acknowledge J. Christensen-Dalsgaard and M. Gabriel for private
communications. Many stimulating and helpful discussions with: G. Alecian,
N. Audard, A.I. Boothroyd, R. Cayrel, D. Cordier, W. Däppen,
J.M. Marques, G. Michaud, J. Montalban, A. Noels, L. Piau, J. Reiter,
S. Turck-Chi\`eze, have brought many improvements. We express our gratitude
to all of them. 
We wish to express our thanks to the anonymous 
referee whose comments and remarks greatly helped to improve the presentation 
of this paper.

\end{acknowledgements}

\end{document}